\documentclass[10pt,twocolumn,aps,pra,amsmath,amssymb,showpacs]{revtex4-1}
\usepackage{bm}
\usepackage{mathrsfs}
\usepackage{graphicx}
\usepackage[usenames,dvipsnames,svgnames,table]{xcolor}
\usepackage{framed}
\usepackage{amsthm}
\usepackage[utf8]{inputenc}

\colorlet{shadecolor}{blue!15}
\usepackage[unicode=true,
            pdfusetitle, 
            bookmarks=true,
            bookmarksnumbered=false,
            bookmarksopen=false,
            breaklinks=true,
            pdfborder={0 0 0},
            backref=false,
            colorlinks=true]{hyperref}
\hypersetup{linkcolor=NavyBlue,urlcolor=NavyBlue,citecolor=NavyBlue}

\usepackage{amsthm}
\newcommand{\propnumber}{} 
\newtheorem*{prop}{Proposition \propnumber}
\newenvironment{propc}[1]
  {\renewcommand{\propnumber}{#1}%
  \begin{prop}}
  {\end{prop}}
  \newtheorem{propo}{Proposition} 
\newtheorem{thm}{Theorem}

 \newtheorem{defn}[thm]{Definition}
 
 \usepackage{tikz-cd}

\newcommand{\cH}{\mathcal{H}}

\begin{document}

\title {Linearity conditions leading to complete positivity}

\author {Iman Sargolzahi}
\email{sargolzahi@neyshabur.ac.ir; sargolzahi@gmail.com}
\affiliation {Department of Physics, University of Neyshabur, Neyshabur, Iran}


\begin{abstract}
The reduced dynamics of an open quantum system $S$, interacting with its environment $E$, is not completely positive, in general. In this paper, we demonstrate that if the two following conditions are satisfied, simultaneously, then the reduced dynamics is completely positive: (1) the reduced dynamics of the system is linear, for  arbitrary system-environment  unitary evolution $U$; and (2) the reduced dynamics of the system is linear, for  arbitrary initial state of the system $\rho_S$.
\end{abstract}

\maketitle

\section{Introduction}

In the axiomatic approach to  quantum operations, as legitimate maps describing the (reduced) dynamics of a quantum system $S$, a quantum operation  $\mathcal{E}_S$ is defined as a linear trace-preserving completely positive map \cite{a1}. At first glance, requiring that  $\mathcal{E}_S$ is linear  seems admissible, since the unitary evolution of a closed quantum system is linear, and we may expect similar property for open quantum systems too.
 In addition, nonlinear evolution may lead to  superluminal signaling \cite{a2}.

But, instead of being trace-preserving \textit{completely positive}, one may expect that  $\mathcal{E}_S$ must  be solely a trace-preserving \textit{positive} map, since the only general requirement seems to be that   $\mathcal{E}_S$ must map density operators to density operators.

It seems that there are two major reasons, for the usual use of  completely positive maps, instead of the positive ones, in quantum information theory \cite{a1}, and in the theory of open quantum systems \cite{a26, a27, a28}: First, there exists a simple operator sum representation, for each  trace-preserving completely positive (CP) map  $\mathcal{E}_S$, as 
\begin{equation}
\label{eq:a2}
\begin{aligned}
\mathcal{E}_S(\rho_{S})=\sum_{i} E_{i}\,\rho_{S}\,E_{i}^{\dagger},\ \ \ \ \ 
  \sum_{i} E_{i}^{\dagger}E_{i}=I_{S}, 
\end{aligned}
\end{equation}
where $E_{i}$ are linear operators and $I_S$ is the identity operator, on the Hilbert space of the system $\cH_S$ \cite{a1}.

Second, in the theory of open quantum systems, it is common to consider the set of  initial states of the system-environment  as $\mathcal{S}=\lbrace\rho_{SE}=\rho_S\otimes \tilde{\omega}_E\rbrace$, where $\rho_S$ is an arbitrary state (density operator) on $\cH_S$ and $\tilde{\omega}_E$ is a fixed state on the  Hilbert space of the environment $\cH_E$ \cite{a26, a27, a28}. Then, for such an initial set $\mathcal{S}$, it is famous that the reduced dynamics of the system is CP, for arbitrary system-environment unitary evolution $U$ \cite{a1}.

The main question of this paper is to investigate whether it is possible to result the CP-ness of the reduced dynamics, from its positivity, or even from the less restrictive condition of its linearity. 

Unlike the reduced dynamics, for which,  in general, its positivity is not equivalent to its CP-ness, there exists an important map for which it is so. This important map is the inverse of the partial trace over the environment, and is called the \textit{assignment map} \cite{a23, a24}. It can be shown that if there exists a positive assignment map, then there exists a CP one too, which results in the CP-ness of the reduced dynamics \cite{a25}.


 As we will see, in  Sec. \ref{sec: D}, only requiring that the reduced dynamics is linear, for arbitrary unitary evolution of the system-environment $U$ and arbitrary initial state of the system $\rho_S$, results in the positivity of the assignment map, and so the CP-ness of the reduced dynamics.

The paper is organized as follows. In the next section, we review some introductory points, on  the reduced dynamics of an open quantum system. The assignment map, and its role in representing the reduced dynamics as a linear map, is introduced in Sec. \ref{sec: C}. Our main results are given in  Sec. \ref{sec: D}, and the paper is ended in  Sec. \ref{sec: E}, with a summary of our results.

\section{Reduced dynamics of an open system}  \label{sec: B}


Let us denote the set of all  linear operators on $\cH_S$ as $\mathcal{L}_S$, and the set of all density operators on   $\cH_S$ as $\mathcal{D}_S$. Now, by a Hermitian map, we mean  a linear trace-preserving map on $\mathcal{L}_S$, which maps each Hermitian operator to a Hermitian operator. 
A Hermitian map is called positive, if it maps each density operator, in $\mathcal{D}_S$, to a density operator. Both, Hermitian maps and positive ones, have operator sum representations as
\begin{equation}
\label{eq:a1}
\begin{aligned}
\Phi_S(\rho_{S})=\sum_{i}e_{i}\,\tilde{E_{i}}\,\rho_{S}\,\tilde{E_{i}}^{\dagger},\quad   \sum_{i}e_{i}\,\tilde{E_{i}}^{\dagger}\tilde{E_{i}}=I_{S}, 
\end{aligned}
\end{equation}
where $\tilde{E_{i}}$ are linear operators  on  $\cH_S$, and $e_{i}$ are real coefficients \cite{a3, a4, a5}.
When all of the coefficients $e_{i}$ in Eq. \eqref{eq:a1} are positive, we can define $E_{i}=\sqrt{e_{i}}\,\tilde{E_{i}}$, and Eq. \eqref{eq:a1} can be rewritten as Eq. \eqref{eq:a2}. Then, the map is called CP. It is also worth noting that the CP-ness of the map $\mathcal{E}_S$, in  Eq. \eqref{eq:a2}, is equivalent to the positivity of the map $ \mathrm{id}_W\otimes\mathcal{E}_S $, where the \textit{witness}  $W$ is an arbitrary (finite
dimensional) quantum system, distinct from the system $S$ (and the environment $E$), and  $\mathrm{id}_W$ is the identity map on    $\mathcal{L}_W$ \cite{a1}.
($\mathcal{L}_W$ is the set of all linear operators on the Hilbert space of the witness $\cH_W$.)

For the open quantum system $S$, interacting with its environment $E$, we can consider the whole system-environment as a closed quantum system, which evolves unitarily as 
\begin{equation}
\label{eq:a11}
 \rho_{SE}^{\prime}=\mathrm{Ad}_{U}(\rho_{SE})\equiv  U \rho_{SE} U^{\dagger},
 \end{equation} 
 where $U$ is a unitary operator, on $\cH_S\otimes\cH_E$. In addition, $\rho_{SE}$ and $\rho_{SE}^{\prime}$ are  initial and  final states of the system-environment, respectively. So, 
the reduced dynamics of the system is given by  
\begin{equation}
\label{eq:a12a}
\rho_{S}^{\prime}=\mathrm{Tr}_{E}(\rho_{SE}^{\prime})=\mathrm{Tr}_{E} \circ \mathrm{Ad}_{U}(\rho_{SE}).
\end{equation} 

In general, the reduced dynamics of the system $S$ cannot be represented by a map  \cite{a11, a5}, i.e., $\rho_{S}^{\prime}$ cannot be given as a function of the initial state of the system $\rho_{S}=\mathrm{Tr}_{E}(\rho_{SE})$, in general. Even if the reduced dynamics of the system can be given by a map, this map is not linear, in general \cite{a12, a13}. And, even if it is linear, it is not (completely) positive, in general, but it is Hermitian \cite{a14}. 
The CP-ness of the reduced dynamics has been proven, only for some restricted sets $\mathcal{S}=\lbrace\rho_{SE}\rbrace$, 
 of  initial states of the system-environment  \cite{a15, a16, a17, a18, a19, a20, a21}.

In the experimentally relevant cases, one usually deals with the factorized initial states of the system-environment, i.e., the set of  initial states of the system-environment, at time $t=0$,  is as 
 $\mathcal{S}=\lbrace\rho_S\otimes \tilde{\omega}_E\rbrace$, where $\rho_S$ is an arbitrary state of the system,  while  $\tilde{\omega}_E$ is a fixed state of the environment \cite{a26, a27, a28}. So, the reduced dynamics is CP, as stated in the Introduction.
But, even in such cases, one may encounter non-CP reduced dynamics, simply by changing the initial time from $t=0$, as illustrated  in the following example.

 Consider the case that the reduced dynamics is given by a  master equation, which is similar to  the Gorini-Kossakowski-Sudarshan-Lindblad one \cite{40b, 41b}, but with a  time-dependent generator 
 $\mathcal{K}_S(t)$,  as 
\begin{equation}
\label{eq:b5}
\begin{aligned}
\frac{d\sigma_{S}}{dt} =\mathcal{K}_S(t)[\sigma_S] \qquad\qquad \qquad\qquad \qquad\qquad\qquad\qquad \quad \\
=-\frac{i}{\hbar} [H(t), \sigma_S] \qquad\qquad \qquad\qquad \qquad\qquad\qquad\quad \ \\
\qquad + \sum_j \gamma_j(t)
  \left[A_j(t)\sigma_S A_j^{\dagger}(t) -\frac{1}{2} 
  \lbrace A_j^{\dagger}(t)A_j(t), \sigma_S 
  \rbrace \right],
\end{aligned}
\end{equation}  
where $\sigma_S=\sigma_S(t) \in \mathcal{D}_S$ is the reduced state of the system $S$, at time $t$. In addition, the (Hermitian) Hamiltonian operator $H(t) \in \mathcal{L}_S$, the Lindblad operators $A_j(t)\in \mathcal{L}_S$, and the real rates $\gamma_j(t)$ are all time-dependent, in general \cite{42b}.
Now, if all $\gamma_j(t)$ are positive, for all $t\geq 0$, then the reduced dynamics is CP-divisible \cite{42b}:
\begin{equation}
\label{eq:b6}
\begin{aligned}
\mathcal{E}_S(t_2, 0)=\mathcal{E}_S(t_2, t_1)\circ \mathcal{E}_S(t_1, 0), 
\end{aligned}
\end{equation}  
where $t_2>t_1>0$, and $\mathcal{E}_S(t, s)$ is a CP map, which maps $\sigma_S(s)$ to $\sigma_S(t)$. But, if, in the \textit{canonical} form of the generator $\mathcal{K}_S(t)$ \cite{43b}, all $\gamma_j(t)$ are positive, only during the time interval $[0, t_1]$, then, we have
\begin{equation}
\label{eq:b7}
\begin{aligned}
\mathcal{E}_S(t_2, 0)=\Phi_S(t_2, t_1)\circ \mathcal{E}_S(t_1, 0), 
\end{aligned}
\end{equation} 
where, though $\mathcal{E}_S(t_1, 0)$ and $\mathcal{E}_S(t_2, 0)$ are CP, but $\Phi_S(t_2, t_1)$, i.e., the Hermitian map which maps $\sigma_S(t_1)$ to $\sigma_S(t_2)$, is non-CP, in general.
So, changing the initial time, from $t=0$ to $t=t_1$, results that the reduced dynamics of the system is given by the non-CP map  $\Phi_S(t, t_1)$, for $t>t_1$.

In addition to  simplicity and  experimental relevance, which were mentioned above and in the Introduction, one can give a rather general discussion, leading to the CP-ness of the reduced dynamics:
always, in addition to the system under study $S$, one can consider another quantum system, the witness $W$, which does not interact with  $S$, and, during the evolution of $S$, it does not evolve. Now, assuming that the evolution of the witness-system is given by a local  map $ \mathrm{id}_W\otimes\mathcal{E}_S $, results in the CP-ness of  $\mathcal{E}_S $.
Note that the initial state of  the witness-system $\rho_{WS}$ can be entangled. 
Now, the CP-ness of $\mathcal{E}_S$, and so the positivity  of  the  $ \mathrm{id}_W\otimes\mathcal{E}_S$, is necessary to ensure that the final state $\rho^{\prime}_{WS}=\mathrm{id}_W\otimes\mathcal{E}_S(\rho_{WS})$ is a valid density operator \cite{a1}.
However, one can find situations in which, though the dynamics of the  witness-system is local (and the reduced state of the witness does not change, during the evolution), it cannot be written as  $ \mathrm{id}_W\otimes\mathcal{E}_S$ (see, e.g. \cite{a9aa}). So, the reduced dynamics of the system $S$  can be non-CP, in general, as we have seen for $\Phi_S(t, t_1)$, in the previous paragraph.


 At the end of this section, we  mention that the utilization of the completely positive maps, for describing the reduced dynamics of the system $S$,  can be extended, at least, through the two following ways. 
First, consider the case that the  set of  initial states of the system-environment  is given by  
 $\mathcal{S}=\lbrace \rho_{SE}= \sum_{\alpha} \tilde{w}_{\alpha} Q_{\alpha} \otimes \tilde{\sigma}_\alpha \rbrace$, where the linear operators  $ Q_{\alpha} \in \mathcal{L}_S$  vary, by changing $\rho_{SE}$, but $\tilde{\sigma}_\alpha $ are fixed density operators on $\cH_E$, and  the (positive)  weights 
   $\tilde{w}_{\alpha}$ are also  fixed. 
Then, the reduced dynamics of the system $S$, in Eq. \eqref{eq:a12a}, for arbitrary system-environment unitary evolution $U$, is given by 
\begin{equation}
\label{eq:bb7}
\begin{aligned}
\rho_S^\prime = \sum_\alpha \tilde{w}_{\alpha} \mathcal{E}_S^{(\alpha)}(Q_\alpha),
\end{aligned}
\end{equation} 
where $\mathcal{E}_S^{(\alpha)}$ is a CP map, depending on $U$ and  $\tilde{\sigma}_\alpha $ \cite{bb1}.
In other words, in this case, the reduced dynamics is given by a set of CP maps $\lbrace\mathcal{E}_S^{(\alpha)}\rbrace$, instead of only one CP map.

Second, consider the case that set of  initial states of the system-environment  is given by 
 $\mathcal{S}=\lbrace \rho_{SE}= \mathcal{E}_S \otimes  \mathrm{id}_E (\tilde{\omega}_{SE}) \rbrace$, where $\tilde{\omega}_{SE}$ is a fixed state on $\cH_S \otimes \cH_E$, $\mathcal{E}_S$ is an arbitrary CP map on $\mathcal{L}_S$, and  $\mathrm{id}_E$ is the identity map on $\mathcal{L}_E$, the set of all linear operators on $\cH_E$. Splitting a quantum experiment into the three steps of preparation, evolution and measurement, choosing the  set $\mathcal{S}$ as above means that  we can only manipulate the system $S$,
 through the CP maps $\mathcal{E}_S$, during the preparation step. 
Now, it can be shown that,   for arbitrary system-environment unitary evolution $U$, the final state of the system $\rho_S^\prime $, in Eq. \eqref{eq:a12a}, can be written as  a completely positive map on (the Choi matrix representation  \cite{a6, a7}  of)  $\mathcal{E}_S$ \cite{bb15, bb2}.
 In other words, in this case, even if $\rho_S^\prime $ cannot be given as a completely positive map on the initial state of the system $\rho_S$, but it can be given by a completely positive map, on the preparation map $\mathcal{E}_S$.

\section{Assignment map}  \label{sec: C}

Consider the set $\mathcal{S}=\lbrace\rho_{SE}\rbrace$ of  initial states of the system-environment. 
The set $\mathcal{S}$ includes all initial $\rho_{SE}$ which are prepared (chosen), through the preparation step of the experiment. Obviously, in general, $\mathcal{S}$ is a subset of $\mathcal{D}$, the set of all density operators on $\cH_S\otimes \cH_E$.

The set of  initial states of the system is given by $\mathcal{S}_S=\mathrm{Tr}_E\mathcal{S}$.
Assuming that the system $S$ is finite dimensional, of dimension $d_S$, only a finite number $m$ of the members of $\mathcal{S}_S$, where the integer $m$ is  $0< m\leq {(d_S)}^2$, are linearly independent. Let us denote this linearly independent set as
  $\mathcal{S}^\prime_S  = \lbrace\rho_{S}^{(1)}, \rho_{S}^{(2)}
  , \ldots\ , \rho_{S}^{(m)}\rbrace$. Therefore, any $\rho_{S}\in \mathcal{S}_S$ can be expanded as
\begin{equation}
\label{eq:a4}
\begin{aligned}
 \rho_{S}=\sum_{i=1}^m a_i \rho_{S}^{(i)},
\end{aligned}
\end{equation}  
   where $a_i$ are real coefficients. Note that $ \rho_{S}$ is a Hermitian operator.  So,  $\sum (a_i-a_i^*) \rho_{S}^{(i)}=0$. 
   Now, since all $\rho_{S}^{(i)} \in \mathcal{S}_S^\prime$ 
   are linearly independent, all   $a_i$ must be real.

In general, there may be more than one state in  $\mathcal{S}$ such that tracing over the environment gives $\rho_{S}^{(i)}$. However, we choose only one of them and denote it as  $\rho_{SE}^{(i)}$. 
Linear independence of   $\rho_{S}^{(i)}\in \mathcal{S}^\prime_S$ results in linear independence of  $\rho_{SE}^{(i)}$.
 We denote this linearly independent set as   $\mathcal{S}^\prime  = \lbrace\rho_{SE}^{(1)}, \rho_{SE}^{(2)}
  , \ldots\ , \rho_{SE}^{(m)}\rbrace$   \cite{b1}. So, each $\rho_{SE}\in \mathcal{S}$, for which $\rho_{S}=\mathrm{Tr}_E(\rho_{SE})$ is expanded in Eq. \eqref{eq:a4}, can be written as 
 \begin{equation}
\label{eq:a5}
\begin{aligned}
 \rho_{SE}=\sum_{i=1}^m a_i \rho_{SE}^{(i)}+Y( \rho_{SE}),
\end{aligned}
\end{equation}  
   where $a_i$ are the same as those  in Eq. \eqref{eq:a4},  and $Y$ is a Hermitian operator, on $\cH_S\otimes\cH_E$, such that $\mathrm{Tr}_E (Y)=0$. In other words, Eq. \eqref{eq:a4} results that   $\rho_{SE}$ and $\sum  a_i \rho_{SE}^{(i)}$ can differ with each other up to a Hermitian operator $Y$, for which $\mathrm{Tr}_E (Y)=0$. In general, $Y$ is a function of $\rho_{SE}$. This dependence is explicitly given in Eq. \eqref{eq:a5}, by writing it as $Y( \rho_{SE})$.
 
\begin{figure}
\begin{center}
\includegraphics[width=8.5 cm]{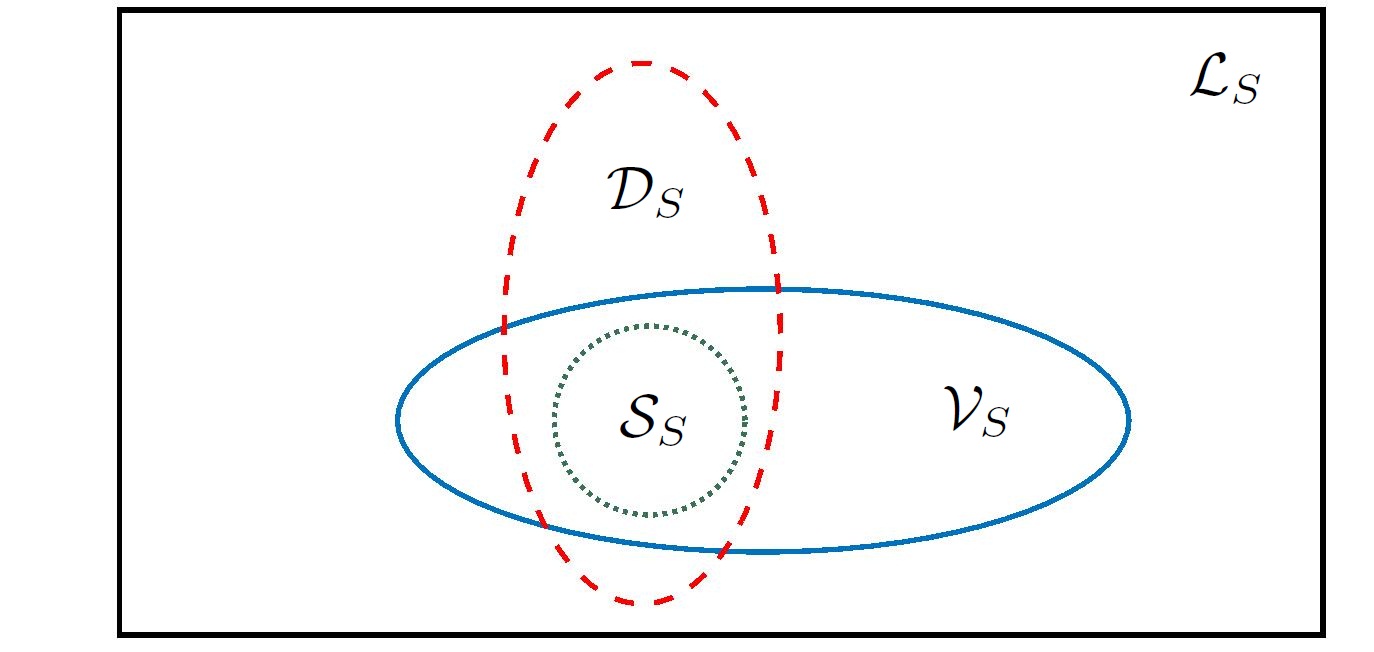}
\end{center}
\caption{The set $\mathcal{S}_S=\mathrm{Tr}_E\mathcal{S}$ (green dotted circle) is the set of  initial states of the system $S$. 
The set $\mathcal{D}_S$ (red dashed ellipse) is the set of all states (density operators) on $\cH_S$. Obviously, $\mathcal{S}_S\subseteq \mathcal{D}_S$.
The subspace $\mathcal{V}_S$ (blue solid ellipse) is defined in Eq. \eqref{eq:a7}, and so, 
$\mathcal{S}_S \subset\mathcal{V}_S$. Finally, $\mathcal{L}_S$ (black solid rectangle) is the set of all linear operators on  $\cH_S$. So,  $\mathcal{D}_S \subset\mathcal{L}_S$ and $\mathcal{V}_S \subseteq\mathcal{L}_S$. When $\mathcal{S}_S = \mathcal{D}_S$, then $\mathcal{V}_S = \mathcal{L}_S$.}
\label{Fig1}
\end{figure}

The subspaces $\mathcal{V}$ and $\mathcal{V}_S$ are defined as \cite{a5}
 \begin{equation}
\label{eq:a6}
\begin{aligned}
\mathcal{V}= \mathrm{Span}_{\mathbb{C}} \  \mathcal{S},  
\end{aligned}
\end{equation}  
and 
\begin{equation}
\label{eq:a7}
\begin{aligned}
\mathcal{V}_S=\mathrm{Tr}_{E} \mathcal{V}=\mathrm{Span}_{\mathbb{C}} \  \mathcal{S}_S=\mathrm{Span}_{\mathbb{C}} \  \mathcal{S}_S^\prime .
\end{aligned}
\end{equation}  
Therefore, each $X \in \mathcal{V}$ can be written as $X=\sum_{l} c_l \, \tau_{SE}^{(l)}$, where 
$\tau_{SE}^{(l)} \in  \mathcal{S}$, and $c_l$ are complex coefficients.
Using Eq. \eqref{eq:a5}, we can expand each $\tau_{SE}^{(l)}$ as  $\tau_{SE}^{(l)}=\sum_{i} a_{li} \rho_{SE}^{(i)}+Y^{(l)}$. So,
\begin{equation}
\label{eq:a8}
\begin{aligned}
 X=\sum_{i=1}^m  \left( \sum_{l}  a_{li} c_l \right) \rho_{SE}^{(i)}+\sum_{l} c_l \,  Y^{(l)} \ \\
 =\sum_{i=1}^m d_i \rho_{SE}^{(i)}+Y(X), \qquad\qquad\qquad
\end{aligned}
\end{equation}  
where $d_i=\sum_{l}  a_{li} c_l $ are complex coefficients, and the linear operator $Y(X)=\sum_{l} c_l \,  Y^{(l)}$ is such that $\mathrm{Tr}_E (Y(X))=0$. Consequently, for each $x\in \mathcal{V}_S$, we have
\begin{equation}
\label{eq:a9}
\begin{aligned}
x =\mathrm{Tr}_E (X) =\sum_{i=1}^m d_i \rho_{S}^{(i)},
\end{aligned}
\end{equation}  
where the  coefficients $d_i$ are the same as those in Eq.  \eqref{eq:a8}. In Fig. \ref{Fig1}, the sets 
$\mathcal{S}_S$ and $\mathcal{D}_S$,  the subspace $\mathcal{V}_S$, and the vector space $\mathcal{L}_S$ are given, in a Venn diagram.

Now, we can define the linear trace-preserving assignment map $ \Lambda_S $, as follows:  first, we define $\Lambda_S(\rho_{S}^{(i)})=\rho_{SE}^{(i)}$. Then, we extend the definition of $ \Lambda_S $, to the whole $\mathcal{V}_S$,  as a linear map. So, 
for any $x\in \mathcal{V}_S$, in Eq. \eqref{eq:a9}, we have
\begin{equation}
\label{eq:a10}
\begin{aligned}
\Lambda_S(x)=\sum_{i=1}^m d_i \Lambda_S(\rho_{S}^{(i)})=\sum_{i=1}^m d_i \rho_{SE}^{(i)}.
\end{aligned}
\end{equation} 
The assignment map $ \Lambda_S $ maps $\mathcal{V}_S$ to (a subspace of)  $\mathcal{V}$, and is  Hermitian, by construction. (When $x$ is a Hermitian operator, all $d_i$, in Eq.  \eqref{eq:a9}, are real. So, $\Lambda_S(x)$ is also a Hermitian operator.) Comparing Eqs. \eqref{eq:a8} and \eqref{eq:a10} shows that  $ \Lambda_S $ does not necessarily map $x$ to $X$, unless $Y(X)=0$.
In addition, note that the assignment map $ \Lambda_S $, in Eq.  \eqref{eq:a10}, is defined on the subspace $\mathcal{V}_S$.
This definition can be extended,  to the whole  $\mathcal{L}_S$, simply, i.e., one can find a  Hermitian map $\Lambda_S^\prime $, on the whole  $\mathcal{L}_S$, such that, for each $x\in \mathcal{V}_S$, it acts as $ \Lambda_S $ \citep{a25}. But, only for each $x\in \mathcal{V}_S$,  not necessarily for arbitrary $ f \in \mathcal{L}_S$, we have $\mathrm{Tr}_{E} \circ \Lambda_S^\prime (x)=\mathrm{Tr}_{E} \circ \Lambda_S (x)= x$. In other words, the \textit{extension} 
  $\Lambda_S^\prime $ of the assignment map $ \Lambda_S $ is \textit{self-consistent} only  on $\mathcal{V}_S$, not
   necessarily on the whole $\mathcal{L}_S$.

\begin{figure}
\begin{center}
\includegraphics[width=8.5 cm]{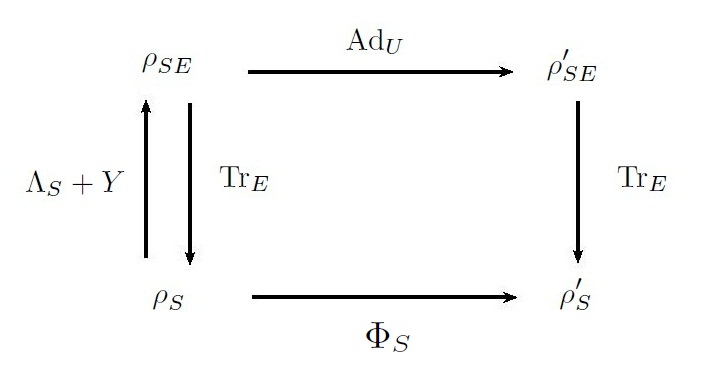}
\end{center}
\caption{The state $\rho_{SE}$ is the initial state of the whole system-environment. The final state of the system-environment $\rho_{SE}^{\prime}$ is given in Eq. \eqref{eq:a11}. Tracing over the environment $E$,  gives the initial state of the system $\rho_{S}=\mathrm{Tr}_{E}(\rho_{SE})$, and its final state  $\rho_{S}^{\prime}=\mathrm{Tr}_{E}(\rho_{SE}^{\prime})$. According to Eqs. \eqref{eq:a4}, \eqref{eq:a5} and \eqref{eq:a10},  $ \Lambda_S (\rho_{S})+Y$ gives $\rho_{SE}$. The  map $\Phi_S$ is defined as $\Phi_S = \mathrm{Tr}_{E} \circ \mathrm{Ad}_{U}  \circ \Lambda_S$. 
According to Eq. \eqref{eq:a12},  $\Phi_S$  gives $\rho_{S}^{\prime}$, if the $U$-consistency condition $\mathrm{Tr}_{E} \circ \mathrm{Ad}_{U}(Y)=0$ is satisfied. Then, rounding the diagram clockwise, from $\rho_{S}$ to $\rho_{S}^{\prime}$, is equivalent to rounding it counterclockwise, through the Hermitian map $\Phi_S$. }
\label{Fig2}
\end{figure}

 Now, using Eqs. \eqref{eq:a12a}, \eqref{eq:a4}, \eqref{eq:a5}  and \eqref{eq:a10}, the reduced dynamics of the system, for each $\rho_{SE} \in \mathcal{V}$, is given by  
\begin{equation}
\label{eq:a12}
\begin{aligned}
\rho_{S}^{\prime}=\mathrm{Tr}_{E} \circ \mathrm{Ad}_{U}(\rho_{SE}) \qquad\qquad \qquad\qquad\quad\quad  \\
 = \sum_{i=1}^m a_i \mathrm{Tr}_{E} \circ \mathrm{Ad}_{U}(\rho_{SE}^{(i)})+\mathrm{Tr}_{E} \circ \mathrm{Ad}_{U}(Y)  \\
\qquad =\mathrm{Tr}_{E} \circ \mathrm{Ad}_{U}  \circ \Lambda_S(\rho_{S})+\mathrm{Tr}_{E} \circ \mathrm{Ad}_{U}(Y)   \quad   \\
= \Phi_S(\rho_{S})+\mathrm{Tr}_{E} \circ \mathrm{Ad}_{U}(Y), \qquad\quad\qquad\quad \
\end{aligned}
\end{equation} 
where $\Phi_S\equiv \mathrm{Tr}_{E} \circ \mathrm{Ad}_{U}  \circ \Lambda_S$. 
The map $\Phi_S$ is a (linear)  Hermitian map on  $\mathcal{V}_S$, since $\mathrm{Tr}_{E}$ and $\mathrm{Ad}_{U}$ are CP \cite{a1}, and the 
assignment map $ \Lambda_S $ is Hermitian on  $\mathcal{V}_S$, as we have seen in Eq. \eqref{eq:a10}.  When $\mathrm{Tr}_{E} \circ \mathrm{Ad}_{U}(Y)=0$, the subspace $\mathcal{V}$ is called  $U$-\textit{consistent} \cite{a5}. 
 The reduced dynamics of the system,  for each $\rho_{SE}\in \mathcal{V}$,   is given by the linear Hermitian trace-preserving map $\Phi_S$,  if and only if  $\mathcal{V}$ is $U$-consistent \cite{a22, a5}.
In Fig. \ref{Fig2}, we represent when the Hermitian  map $\Phi_S$ gives the reduced dynamics of the system, in a commutative diagram.
It is also worth noting that,
  in the theory of open quantum systems, one usually  approximates  the reduced dynamics as a linear map, utilizing some simplifying assumptions (about $\mathcal{V}$)  \cite{a26, a27, a28, 28a}.

 CP-ness of $\mathrm{Tr}_{E}$ and $\mathrm{Ad}_{U}$ results that only the assignment map $ \Lambda_S $  determines whether $\Phi_S$ is CP or not.
If $ \Lambda_S $  is  Hermitian, then $\Phi_S$ can be either Hermitian, positive or CP. But, when the extension $\Lambda_S^\prime $  of the assignment map $ \Lambda_S $ is positive, then $\Phi_S$ is necessarily CP \cite{a25}.

We end this section, with the following point. Assuming  unitary  dynamics for the  whole system-environment, the (non)linearity of the reduced dynamics is only a consequence of $U$-(in)consistency of the subspace $\mathcal{V}$. In other words,   it is only a consequence of  how we choose (construct) the initial set $\mathcal{S}$, and there is no fundamental reasoning behind it \cite{a22}. In addition, as discussed
 in Ref. \cite{a22}, non-linearity of the reduced dynamics does not lead to superluminal signaling.

\section{Main result}  \label{sec: D}

Assume that the reduced dynamics of the system, for each $\rho_S \in \mathcal{S}_S$ is given by a \textit{dynamical map} $\Psi_S$, i.e.,  the final state  $\rho_S^\prime$, in Eq. \eqref{eq:a12a}, is given by $\Psi_S(\rho_S)$. 
As discussed in the Introduction, in the axiomatic approach to  quantum operations, postulating that the dynamical map $\Psi_S$  is linear seems more natural than postulating it as a CP map. In addition, it can be shown simply \cite{a22} that  when the map $\Psi_S$ is linear, on the subspace $\mathcal{V}_S$, then it is equal to $\Phi_S$, in  Eq. \eqref{eq:a12}. Now, we ask,
 under what circumstances, does only requiring that  $\Psi_S$ is linear (and so is equal to $\Phi_S$, in  Eq. \eqref{eq:a12}) result that it is also  CP? Such circumstances are given in the following Proposition.

\begin{propo}
\label{pro:1}
Requiring that 
  the reduced dynamics of the system, for each  $\rho_S \in \mathcal{D}_S$, and for arbitrary system-environment
unitary evolution $U$, is a linear function of  $\rho_S$, results in the CP-ness of
  the assignment map $\Lambda_S $. Thus, the reduced dynamics of the system $S$ is CP, as Eq. \eqref{eq:a2}.
\end{propo}
\textit{Proof.}  
First, we require that the reduced dynamics of the system, for arbitrary system-environment
unitary evolution $U$, is  linear. So, the reduced dynamics is given by  the  map $\Phi_S$, in  Eq. \eqref{eq:a12}, for arbitrary $U$ \cite{a22}.
 In other words, the subspace $\mathcal{V}$, in Eq. \eqref{eq:a6}, is $U$-consistent, for arbitrary $U$. This results in the one to one correspondence between the subspaces  $\mathcal{V}$ and  $\mathcal{V}_S=\mathrm{Tr}_{E} \mathcal{V}$ \cite{a5}. Hence, for each $X, Z \in \mathcal{V}$, $\mathrm{Tr}_{E}(X)=\mathrm{Tr}_{E}(Z)$ if and only if $X=Z$. It indicates that $Y(\rho_{SE})$, in Eq. \eqref{eq:a5}, and so $Y(X)$, in Eq. \eqref{eq:a8}, are zero.
 Therefore, $\Lambda_S (\rho_S)=\rho_{SE}$ and $\Lambda_S (x)=X$, where the linear assignment map $\Lambda_S $  is defined in Eq. \eqref{eq:a10}, and $\rho_S$, $\rho_{SE}$, $X$ and $x$ are given in Eqs. 
\eqref{eq:a4}, \eqref{eq:a5}, \eqref{eq:a8} and \eqref{eq:a9}, respectively.

Second, we require that  the reduced dynamics of the system is linear, for arbitrary initial state of the system $\rho_S \in \mathcal{D}_S$. This means that we choose the set of initial states of the system-environment  $\mathcal{S}$ such that 
 $\mathcal{S}_S=\mathcal{D}_S$.  
 Therefore, since one can find $(d_S)^2$ linearly independent states in $\mathcal{D}_S$ (see, e.g., \cite{26b}),  
we have $\mathcal{V}_S=\mathrm{Span}_{\mathbb{C}} \  \mathcal{D}_S=\mathcal{L}_S$.

Note that we want to find the conditions which ensure the positivity of (the extension of) the assignment map $\Lambda_S $  in Eq. \eqref{eq:a10}. Requiring that, for a given $U$, the reduced dynamics is linear,  for arbitrary initial state $\rho_S \in \mathcal{D}_S$,  results that $\mathcal{S}_S=\mathcal{D}_S$ (and so $\Lambda_S^\prime=\Lambda_S $, since $\mathcal{V}_S=\mathcal{L}_S$) and $\mathrm{Tr}_{E} \circ \mathrm{Ad}_{U}(Y)=0$, where $Y$ is given in Eq. \eqref{eq:a5}. But, it does not necessitate that $Y=0$.
So, the assignment map $\Lambda_S $, which maps $\rho_S$, in Eq. \eqref{eq:a4}, to $Z=\sum_{i=1}^m a_i \rho_{SE}^{(i)}$, is not necessarily positive, since $Z$ is  not necessarily a positive operator.
But, if  we add the first requirement too, which ensures that $Y=0$, then we conclude that $\Lambda_S =\Lambda_S^\prime$ is positive.

On the other hand, only assuming the first requirement, though results in the positivity of $\Lambda_S $ on  $\mathcal{S}_S$, but it does not necessarily lead to  the positivity of the extension $\Lambda_S^\prime$ of the assignment map $\Lambda_S $,  on   the whole $\mathcal{D}_S$ ($\mathcal{L}_S$). But, if we add the second requirement too, which states that $\mathcal{S}_S=\mathcal{D}_S$, we ensure that $\Lambda_S^\prime=\Lambda_S $ is positive, on the whole $\mathcal{D}_S$ ($\mathcal{L}_S$).

Consequently, assuming that  both the first and the second requirements are satisfied \textit{simultaneously}, results that $\Lambda_S^\prime=\Lambda_S $ is positive, on the whole $\mathcal{D}_S$. Now, it has been shown that when there is a positive extension $\Lambda_S^\prime$ of the assignment map $\Lambda_S $,  on the whole $\mathcal{D}_S$ ($\mathcal{L}_S$), then there exists a CP assignment map $\Lambda_S^{(CP)}$ too \cite{a25}. In fact, in this case, where $\mathcal{S}_S=\mathcal{D}_S$ and so $\Lambda_S^\prime=\Lambda_S $, and, in addition, there is a one to one correspondence between the subspaces  $\mathcal{V}$ and  $\mathcal{V}_S$, there is a unique way to define (the extension of) the assignment map. So,
the CP assignment map $\Lambda_S^{(CP)}$ is the same as our positive  $\Lambda_S = \Lambda_S^\prime$, with 
 the explicit form 
\begin{equation}
\label{eq:a13}
\begin{aligned}
\Lambda_S(\rho_{S})=\Lambda_S^{(CP)}(\rho_{S})=\rho_{S}\otimes \tilde{\omega}_E,
\end{aligned}
\end{equation} 
where $\tilde{\omega}_E$ is a fixed state on $\cH_E$ \cite{a23, a4, a25}. This fact that $\tilde{\omega}_E$ is a fixed state is a consequence of assuming that  the assignment map is a self-consistent positive map, on the whole $\mathcal{D}_S$ ($\mathcal{L}_S$) \cite{a23, a4, a25}.
The assignment map $\Lambda_S^{(CP)}$, given in Eq. \eqref{eq:a13}, is, in fact, the famous Pechukas’s  one, first introduced in Ref. \cite{a23}. Finally, the CP-ness of $\Lambda_S^{(CP)}$ leads to the CP-ness of  the reduced dynamics  $\Phi_S=\mathrm{Tr}_{E} \circ \mathrm{Ad}_{U}  \circ \Lambda_S= \mathrm{Tr}_{E} \circ \mathrm{Ad}_{U}  \circ \Lambda_S^{(CP)}$. $\qquad\qquad\qquad\qquad\qquad \blacksquare$

In the axiomatic approach to  quantum operations, it is more appropriate to postulate that the  dynamical map $\Psi_S$ is \textit{convex-linear}, instead of considering it  linear. A convex-linear map is defined as follows.
\begin{defn}
 When $\Psi_S$ is  convex-linear, on $\mathcal{D}_S$, then we have  $\Psi_S \left( p \rho_{S}+ (1-p) \tau_{S}\right)=p\Psi_S(\rho_{S})+(1-p)\Psi_S(\tau_{S})$, where $\rho_{S}, \ \tau_{S} \in \mathcal{D}_S$ and  $0\leq p\leq 1$.
\end{defn}
In the following Proposition, we refer to the convexity of the set $\mathcal{S}_S$. This property is defined as below.
\begin{defn}
  When $\mathcal{S}_S$ is convex, if $\rho_{S}, \tau_{S}\in \mathcal{S}_S$, then, also, $\omega_{S}= p \rho_{S}+ (1-p) \tau_{S}\in \mathcal{S}_S$, where $0\leq p\leq 1$. 
\end{defn}

In   Proposition \ref{pro:1}, we have seen that requiring  the reduced dynamics of the system $S$  is linear, leads to its CP-ness.
Now, we want to go further and show that  requiring  the reduced dynamics  is convex-linear,  results in the CP-ness of the reduced dynamics too.
\begin{propc}{$\bf{ 1^{\prime}}$}
\label{pro:1a}
Requiring that 
  the reduced dynamics of the system, for each  $\rho_S \in \mathcal{D}_S$, and for arbitrary system-environment
unitary evolution $U$,  is a convex-linear   function of  $\rho_S$,  results in the  CP-ness of the assignment map $\Lambda_S $, as Eq. \eqref{eq:a13}. Thus, the reduced dynamics is CP, as Eq. \eqref{eq:a2}, for arbitrary $U$ and arbitrary   $\rho_S \in \mathcal{D}_S$.
\end{propc}

\textit{Proof.} Since, as before, we have $\mathcal{S}_S=\mathcal{D}_S$, the set  $\mathcal{S}_S$ is convex. Thus, we can show that the  convex-linearity of the reduced dynamics results in its linearity, following a similar procedure as Ref. \cite{a22}.

 Note that some of the real coefficients $a_i$, in Eq. \eqref{eq:a4}, are positive, and the others are negative. Let us denote the positive ones as $a_i^{(+)}$, and the negative ones as $a_i^{(-)}$. So, from  Eq. \eqref{eq:a4}, we have
 \begin{equation}
\label{eq:a14}
\begin{aligned}
 \rho_{S}+\sum_{i} \vert a_i^{(-)} \vert \rho_{S}^{(i)} =\sum_{i}  a_i^{(+)}  \rho_{S}^{(i)}.
\end{aligned}
\end{equation} 
Tracing from both sides, we have $1+\sum_{i} \vert a_i^{(-)} \vert =\sum_{i} a_i^{(+)}  \equiv b$.
Dividing both sides of Eq.  \eqref{eq:a14} into $b$ results in
\begin{equation}
\label{eq:a15}
\begin{aligned}
\frac{1}{b} \left( \rho_{S}+\sum_{i} \vert a_i^{(-)} \vert \rho_{S}^{(i)} \right)=\frac{1}{b} \left( \sum_{i}  a_i^{(+)}  \rho_{S}^{(i)} \right)\equiv \omega_S,  
\end{aligned}
\end{equation}
where $\omega_S \in \mathcal{D}_S=\mathcal{S}_S$
 Therefore, assuming that $\Psi_S$ is convex-linear, on $\mathcal{S}_S$, we have
\begin{equation}
\label{eq:a16}
\begin{aligned}
\Psi_S(\omega_S)=\Psi_S\left(\frac{1}{b} ( \rho_{S}+\sum_{i} \vert a_i^{(-)} \vert \rho_{S}^{(i)})\right) \ \\
=\Psi_S\left(\frac{1}{b} (\sum_{i} a_i^{(+)} \rho_{S}^{(i)} )\right) \qquad\quad  \\
\Rightarrow \quad \frac{1}{b} \left(\Psi_S(\rho_{S})+\sum_{i} \vert a_i^{(-)} \vert  \Psi_S(\rho_{S}^{(i)}) \right)  \\
=  \frac{1}{b}\left(\sum_{i}  a_i^{(+)}   \Psi_S(\rho_{S}^{(i)}) \right), \qquad\quad
\end{aligned}
\end{equation}
which leads to
\begin{equation}
\label{eq:a17}
\begin{aligned}
\Psi_S( \rho_{S})=\sum_{i=1}^m a_i \Psi_S(\rho_{S}^{(i)}).
\end{aligned}
\end{equation}
So, noting Eq. \eqref{eq:a4}, we conclude that $\Psi_S$ is linear. Hence, if $\Psi_S$  is convex-linear, for arbitrary $U$ and arbitrary   $\rho_S \in \mathcal{D}_S$, then it is also linear, for arbitrary $U$ and arbitrary    $\rho_S \in \mathcal{D}_S$. Now, Proposition \ref{pro:1} shows that the assignment map $\Lambda_S $ is CP, as Eq. \eqref{eq:a13}, and so the reduced dynamics of the system $\Psi_S=\Phi_S$ is  also   CP. $\qquad\qquad\qquad\qquad\quad \blacksquare$


\section{Summary} \label{sec: E}

Requiring that the reduced dynamics of the system $S$, interacting with its environment $E$, is (convex) linear means that (1) the reduced dynamics is (convex) linear, for arbitrary system-environment evolution $U$, and (2) the reduced dynamics is (convex) linear, for arbitrary initial state of the system $\rho_S \in \mathcal{D}_S$. 

In Proposition \ref{pro:1} (\propnumber{$ 1^{\prime}$}), it has been shown that the above  requirement  results in the CP-ness of the reduced dynamics. So, in the axiomatic approach to  quantum operations, there is no need to consider the CP-ness as a distinct postulate. It is only a consequence of (convex) linearity.

In addition, when the reduced dynamics is (convex) linear, for arbitrary $U$ and arbitrary $\rho_S$, then
the set of  initial states of the system-environment is as $\mathcal{S}=\lbrace\rho_S\otimes \tilde{\omega}_E\rbrace$, where $\rho_S$ is an arbitrary state of the system, and $\tilde{\omega}_E$ is a fixed state  of the environment. In other words, under such circumstances, the assignment map is as the Pechukas's one \cite{a23}, given in Eq. \eqref{eq:a13}.

%


\subsection*{Acknowledgments}

I would like to thank the  anonymous referees for their useful comments.


\begin{thebibliography}{1}
\bibitem{b1} If one $ \rho_{SE}^{(i)}\in \mathcal{S}^\prime$ is linearly dependent of the others, then it can be written as a linear combination of the others. So, tracing over the environment, $\rho_{S}^{(i)}=\mathrm{Tr}_E(\rho_{SE}^{(i)})$ can be expanded as a linear combination of the other members of $\mathcal{S}^\prime_S$, which is in contradiction to  the assumption of the linear independence of all members of $\mathcal{S}^\prime_S$.
\end{thebibliography}
\end{document}